**Flexible BiSeI/NiO-based X-ray synapses bridging the functions of detection and memory**

Qiao Wang, Pengfei Li, Yushou Song, Jalu Li, Haiying Xiao, Yuqing Wang, Guoliang Ma, Hsu-Sheng Tsai, Ping-An Hu

X-ray synapses can provide machine vision and prescreening of the targets, which significantly reduces labor cost and error rate, especially in medical areas. A flexible X-ray detector can reduce the image distortion in the scenarios where curved surfaces are required and increase the accuracy, such as in multi-slice CT. Here, we demonstrate a flexible X-ray synapse by utilizing a quasi-one-dimensional Bi-based plastic semiconductor (BiSeI), which uniquely combines a high atomic number ($Z_{Bi}$ = 83), excellent carrier transport ($\mu\tau = 1.7 \times 10^{-4}$ $cm^2$ $V^{-1}$), and mechanical flexibility, as the detection material. The plasticity mechanism originates from atomic-chain kinking and sliding. The heterojunction of single-crystal BiSeI and polycrystalline NiO thin films show excellent synaptic functions, including paired-pulse facilitation and dynamic transition from short- to long-term memory, without reliance on von Neumann architectures. The as-made X-ray synapse device achieves an ultrahigh sensitivity (~$10^6$ μC $Gy_{air}^{-1}$ $cm^{-2}$) and an ultralow detection limit (34 nGy $s^{-1}$), and a pJ-level energy consumption, rivaling rigid perovskite single crystals while maintaining robust performance on flexible substrates. The integration with convolutional neural network advances machine vision platforms for orthopedic diseases screening and diagnosis with an accuracy as high as 90%.

## 1. Introduction

X-ray detection is a key technology in medical imaging, nondestructive evaluation, materials synthesis, geological exploration, and paleontological sciences.[1] In general, their data processing follows the von Neumann architecture, with separate sensing and memory divisions, leading to bandwidth bottlenecks and high energy consumptions as frame rates and pixel counts rise.[2,3] These deficits are especially disadvantageous in latency-critical tasks, such as real-time medical imaging, robotic surgery, telesurgery, machine vision screening, and tele-operated nuclear contamination mapping.[4,5] Optoelectronic synapses offer a fundamentally different route, in which the detection, memory and even computation capabilities are integrated in a single element. X-ray synapses can perform in-sensor preprocessing like paired-pulse facilitation (PPF), short-to-long-term memory transition (STM/LTM transition), compressing and filtering data at the point of acquisition. This architecture reduces the data traffic between memory and processor, thereby lowering latency and energy consumption. Consequently, it empowers these devices to tackle machine vision tasks that are challenging for conventional detectors to achieve with the same efficiency. In principle, such synaptic/neuromorphic sensing directly addresses the throughput and power challenges of next-generation X-ray systems.[6]

However, most X-ray synapses reported to date are rigid and flat,[3,7,8] struggling to meet the demands for specialized surface conformation and extreme viewing angles, which are frequently required in cutting-edge applications (e.g., 3D imaging[9,10], 4π radiation metrology[11], multislice CT[12]). The angle-limitation, image distortion and non-uniform response demand complex optics and heavy post-processing.[13] Flexible X-ray optoelectronic synapses provides an efficient solution potentially. Current technical pathways for flexible, synaptic radiation detection face considerable challenges mainly owing to the intrinsic material properties. While organic semiconductors can be engineered with mechanical flexibility, they suffer from small absorption coefficient and low mobilities, which limit the sensitivity. For inorganic polycrystalline semiconductor thin films, two inventible trade-offs limit their application. First, the memory mechanism is based on the crystal imperfections such as grain boundaries and defects, which conversely reduce the carrier mobilities, comprising the detection efficiency.[14,15] Second, pursuing mechanical flexibility by reducing film thickness comes at the cost of decreased X-ray absorption, leading to low photoresponsivity.[16] Therefore, the development of flexible X-ray optoelectronic synapses still faces formidable hurdles.

Herein, we report a quasi-one-dimensional (quasi-1D) plastic inorganic semiconductor BiSeI that possesses excellent mechanical and electrical properties. The bulk BiSeI single crystal exhibits large $Z$ value, large mobility−lifetime ($\mu\tau$) product, good mechanical robustness. Such an inorganic van der Waals material without dangling bonds at surface enables the integration with the memory-functional material. A flexible X-ray optoelectronic synapse based on the heterojunction of BiSeI and NiO-$V_{Ni}$ ($V_{Ni}$: nickel vacancies) is developed. The heterojunction combines the key merits from the constituent materials, including flexibility, large X-ray absorption coefficient, large interception volume, superior charge transport properties, and inherent memory functionality. This combination of materials effectively mitigates the intrinsic limitations of monolithic material and preserves their respective advantages. It exhibits advanced synaptic features including the PPF, STM/LTM transition, and stability on flexible substrates at low-power consumption, to support machine vision platforms in orthopedic diseases screening and diagnosis, thereby aiming to reduce labor cost and misdiagnosis rate, particularly in primary care settings or regions with limited medical resources.

## 2. Results and Discussion

### 2.1. BiSeI plastic van der Waals semiconductor

BiSeI, a quasi-1D chain material, is an N-type semiconductor when grown by chemical transport reactions with an orthorhombic crystal structure, belonging to the Pnma space group.[17] The crystal structure consists of a double-chain $[(BiSeI)_\infty]_2$ aligning along the a-axis as shown in **Figure 1**a. The polar covalent bonds link the intrachain atoms while the interaction between the double chains is primarily the weak van der Waals force. This distinctive quasi-1D structure causes BiSeI to preferably grow in one direction, forming a wire- or rod-shaped crystal.

In this study, the synthesis method based on the previous report[17] is modified to suit our experimental setup. Single crystals of BiSeI are synthesized using a chemical vapor transport (CVT) method, where the iodine ($I_2$) serves as the transport agent. Stoichiometric amounts of Bi, Se, and I were used in the ratio of $N_{Bi}$: $N_{Se}$: $N_I$ = 1:1:1.03, where the 3%-excess I compensated for its volatilization during the synthesis. The total raw materials of 10 g were loaded into a quartz tube, which was then vacuum sealed at 1000 Pa. The quartz tube, whose length is 20 cm and diameter is 3 cm, was heated at 300 °C for 24 hours to fully melt and mix the raw materials. The quartz tube was subsequently placed in the centre of a horizontal tube furnace with dual heating zone, where the low-temperature zone was set at 485 °C and the high-temperature zone containing raw materials was set at 505 °C. The temperature gradient across a spatial region of 10 cm between the two zones was maintained for 15 days. Then, the furnace was gradually cooled down to room temperature, resulting in the growth of black stick-like single crystals in the low-temperature zone. The energy dispersive spectroscopy (EDS) analysis (Figure S1a) clearly shows the characteristic peaks of Bi, Se, and I, with an atomic ratio of approximately 1:1:1. The EDS elemental maps (Figure 1d) confirm that the three elements are evenly distributed throughout the crystal, indicating the high-quality synthesized material without phase separation. The X-ray diffraction (XRD) pattern shown in Figure S1b reveals sharp diffraction peaks at 22.1°, 32.0°, 35.3°, and 45.1°, corresponding to the (210), (310), (320), and (112) crystal planes of BiSeI, respectively. The sharpness of these XRD peaks further indicates high crystallinity. The Raman spectrum (Figure S1c) also displays the characteristic peaks of BiSeI single crystal, testifying the successful synthesis of BiSeI.

As shown in Figure 1b, the BiSeI single crystal exhibits excellent mechanical properties as it is subjected to stress and can be wound around a toothpick like a metal wire without fractures. High angle annular dark-field (HAADF) image (Figure 1c) of bent crystals shows no fractures, establishing their intrinsic plasticity. To the best of our knowledge, this is the first report of a plastic inorganic semiconductor with a quasi-1D or 1D molecular structure. Unlike the plastic 2D/3D semiconductors, the plastic quasi-1D/1D inorganic ones with superior mechanical properties are underexplored. The unique quasi-1D structure provides a theoretical framework for directional high-mobility carrier transport,[18,19] offering a structure advantage for X-ray detection.

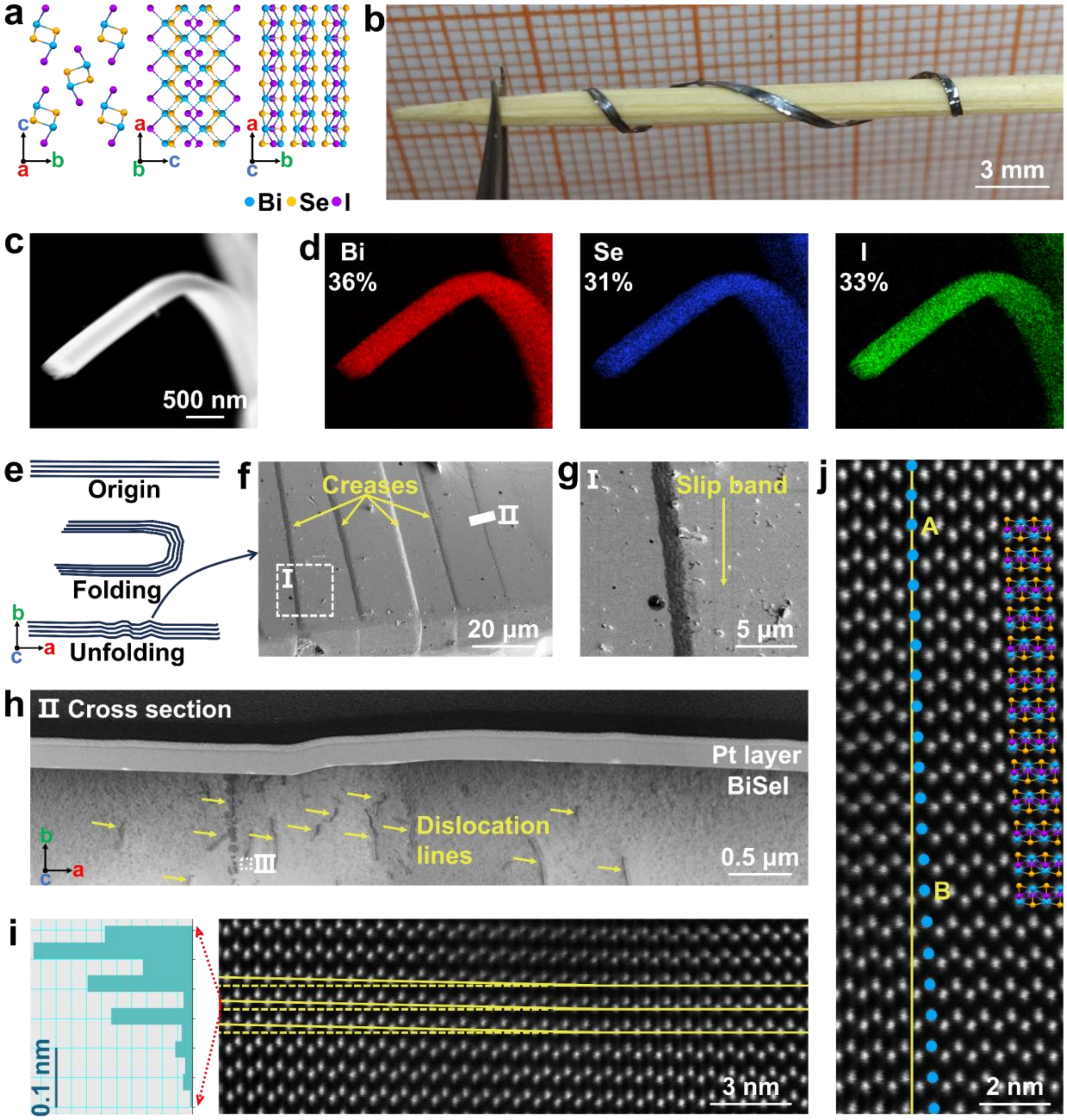

**Figure 1.** Multiscale studies of mechanical plasticity in BiSeI. a) The crystal structure of BiSeI viewed along a, b and c axis, respectively. b) Optical photograph of a BiSeI single crystal morphed into a spring-like shape without breaking. The diameter of the toothpick is 1.5 mm. c) High angle annular dark-field (HAADF) image of bent BiSeI. d) Energy dispersive spectroscopy (EDS) elemental maps of Bi, Se, and I. e) The schematic diagram of the BiSeI crystal undergoing U-shaped folding and recovery under stress along the b-axis. f) Scanning electron microscopy (SEM) image showing the morphology of the (010) plane after deformation. g) Magnified SEM image of the (010) plane revealing step-like creases and parallel stripe-like slip bands. h) Transmission electron microscopy bright-field (TEM-BF) image of the crease profiles showing dislocation lines. i and j) High-angle annular dark-field scanning transmission electron microscopy (HAADF-STEM) images of the BiSeI atomic arrangement in the [001] zone axis projection, showing the kinking (i) and sliding (j) of atomic chains, respectively.

To reveal the plasticity mechanisms, the deformed BiSeI crystals were investigated at multiple scales by various electron microscopies. According to Figure 1e, the BiSeI undergoes U-shaped folding deformation under stress along the b-axis and subsequently recovers to its initial unfolded state. Notably, after one cycle of deformation, step-like creases appear on the (010) crystal plane of BiSeI as shown by the scanning electron microscopy (SEM, Figure 1f), yet the crystal remains intact and crack-free. Parallel stripe-like slip bands are observed on the (010) crystal plane (indicated by the yellow arrows in Figure 1g). The transmission electron microscopy bright-field (TEM-BF) image (Figure 1h) reveals no fractures or interchain cleavage at the crease profiles, with the abundant dislocation lines providing definitive evidence of the exceptional plastic deformation capability of BiSeI.[20]

Having established the role of plastic deformation mediated by dislocations, we employed high-angle annular dark-field scanning transmission electron microscopy (HAADF-STEM) to reveal its atomic-scale structural origin of the plasticity. Since the intensity of individual atomic columns is approximately proportional to $Z^{1.7}$ (where $Z$ is the atomic number), the Se element ($Z_{Se}$ = 34), being lighter than both Bi ($Z_{Bi}$ = 83) and I ($Z_{I}$ = 53), appears nearly invisible along the [001] zone axis.[21] The HAADF-STEM image in Figure S2a exhibits contrast blurring in the core deformation zone compared to the undeformed matrix. This blurring was systematically analyzed by investigating the kinking and sliding of BiSeI atomic chains. As indicated by the yellow solid lines in Figure S2a, kinking causes the BiSeI atomic chains to deviate from their original positions (yellow dashed lines), leading to substantial lattice distortion in the core deformation zone. A local magnified image (Figure 1i) of Figure S2a reveals that the lattice vertical displacement reaches up to 0.3 nm within a lateral distance of ~12 nm. Applying geometric phase analysis (GPA) to Figure S2a further quantifies the distribution of lattice strains during kinking.[22] The $(1\bar{2}0)$ and (120) spots were chosen for GPA and masks were placed around these spots to isolate them, as shown in the inserted Fourier transform (Inset of Figure S2b). The x-direction strain distribution map ($\varepsilon_{xx}$, Figure S2b), indicates that the core deformation zone experiences strain of up to ±20%. The positive and negative values correspond to tensile and compressive stresses, respectively. Additionally, an inverse Fourier transform filtered image of the core deformation zone (Figure S2c) reveals the presence of edge dislocations. These dislocations are characterized by an extra half-plane of atoms, which induces compressive stress on the atoms located above the dislocation line and tensile stress on the atoms below. Therefore, the kinking of atomic chains, which generates dislocations and the associated severe lattice distortions, is a key atomic-scale factor behind the macroscopic plastic behavior of BiSeI. Besides kinking, plastic deformation is also facilitated by the sliding of adjacent atomic chains along the van der Waals interfaces, a process aided by the low energy barrier inherent to such weakly bonded planes.[23] Direct atomic-scale evidence for this sliding mechanism is provided in Figure 1j (a partial magnification of Figure S2a), which captures the coordinated movement of multiple BiSeI chains. The atom from A to B in Figure 1j illustrates a slip of half a unit cell along the a-axis, demonstrating that sustained plastic sliding requires the collective motion of multiple chains. In summary, the synergistic interplay between dislocation-mediated kinking and coordinated chain sliding constitutes the fundamental mechanism governing macroscopic plasticity in BiSeI.

As the mechanism of the plasticity of BiSeI had been elucidated, we systematically studied its mechanical properties. A three-point bending test with a span of 10 mm as shown in **Figure 2**a was performed on a BiSeI single crystal sample with the dimensions of 15×3×0.25 $mm^3$. According to the yield strength (flexural strength) formula (ASTM Standard Test Method: D 790):

$$\delta_f = \frac{3PL}{2bd^2}, \tag{1}$$

where $\delta_f$ is the bending stress, $P$ is the maximum load, $L$ is the span, $b$ is the sample width, and $d$ is the sample thickness. The calculated yield strength of BiSeI is 16.37 MPa, which is two orders of magnitude smaller than that of polycrystalline copper (Cu, 1032 MPa). The low yield strength indicates that BiSeI single crystals can undergo permanent plastic deformation under relatively low bending stresses. The inset in Figure 2a is the optical photograph of the BiSeI sample after the bending test. Despite substantial bending, the BiSeI sample maintains its integrity without visible fractures, directly demonstrating its excellent plastic deformability rather than brittle behavior.

To further probe its mechanical properties at the microscale, nanoindentation tests were conducted, which also demonstrate that BiSeI is unusually soft. In comparison with polycrystalline Cu (Figure S3a, yellow curve), a typical plastic material, the force-displacement curve of BiSeI (Figure S3a, blue curve) is also smooth and rounded without noticeable pop-in effect.[24] To ensure the reliability of the results, the nanoindentation tests were conducted on BiSeI six times. The average Young's modulus (Figure S3b) of BiSeI is approximately 5.78 GPa, significantly smaller than the 145 GPa of polycrystalline Cu. The hardness (Figure S3c) of BiSeI single crystals is lower than the polycrystalline Cu, around 0.889 GPa. As shown in Figure S4, the SEM was used to observe the materials after indentation. The indenter caused no cracks to BiSeI and Cu, unlike that to Si, confirming that the BiSeI exhibits substantial capability of plastic deformation.

In contrast, the $Bi_2Se_3$ composed of similar elements is a brittle 2D layered material,[25] which would be fractured layer-by-layer (Inset of Figure S5b) during the nanoindentation test, leading to a pronounced pop-in effect in the force-displacement curve (Figure S5a). The introduction of iodine (I) causes the material to transform from a 2D structure into a quasi-1D structure, endowing it with more van der Waals interfaces for slip. In addition, the reduction in structure dimension of materials typically alters the physical properties[26], such as Young's modulus, which is supported by the indentation test. The average Young's modulus (Figure S5b) of the 2D $Bi_2Se_3$ is 36.5 GPa, while it is only 6.38 GPa of the quasi-1D BiSeI. This reduction in the Young's modulus reflects a decrease in material rigidity, making it more deformable. The low Young's modulus of BiSeI infers the superiority of quasi-1D or 1D structure for the design of new plastic inorganic materials.

BiSeI exhibits qualities that make it a promising candidate for high-performance flexible X-ray sensors. Beyond meeting essential mechanical prerequisites for flexible devices, this material exhibits superior semiconducting properties critical for X-ray detection, including a large absorption coefficient ($\alpha$). This coefficient depends on the incident energy of photons ($E$) and the atomic number ($Z$ value) of absorbing materials. In general, the $\alpha$ decreases as the $E$ increases, while it sharply increases with the $Z$ value of materials. This relationship could be described by the following equation [27]:

$$\alpha \propto \rho \frac{Z^4}{E^3}, \tag{2}$$

where $\alpha$ is the absorption coefficient, $Z$ is the atomic number, $\rho$ is the mass density, and $E$ represents the X-ray photon energy. Hence, high-$Z$ materials are often preferred for the radiation detectors to achieve high X/γ detection efficiency. It is worth emphasizing that the BiSeI has the highest $Z$ value among the plastic inorganic semiconductors reported to date (Figure 2b), implying its large absorption coefficient. Due to its high atomic number, the absorption coefficient of BiSeI (Figure S6a) is significantly higher than those of commercial materials such as Si, Ge, and CdZnTe at the photon energy above 0.1 MeV. At the photon energy of 0.1 MeV, the required thicknesses to achieve the X/γ ray attenuation of 99.9% is 118 cm for Si, 3.38 cm for Ge, and 0.944 cm for CdZnTe, whereas it is only 0.322 cm for BiSeI (Figure S6b). This indicates that the radiation detectors based on BiSeI should be more efficient as well as more lightweight, making them ideal for the application of portable devices. Furthermore, the smaller and lighter detectors would be advantageous in terms of energy consumption. These features demonstrate the potential of BiSeI for portable nuclear physics research, miniaturized and flexible medical devices, and aerospace applications.

In addition to a high *Z* value, a suitable radiation detection material must exhibit a high $\mu\tau$ product. The $\mu\tau$ product is a key figure of merit for the efficient collection of radiation-induced carriers and can be evaluated using the single-carrier Hecht model[28–30]:

$$\mathrm{CCE} = \frac{\mu\tau V}{d^2}\left[1 - e^{\left(\frac{-d^2}{\mu\tau V}\right)}\right], \tag{3}$$

where CCE is the charge collection efficiency, *V* is the applied bias, *d* is the electrode distance (250 μm), and $\mu\tau$ is the product of carrier mobility ($\mu$) and lifetime ($\tau$). The charge collection efficiency is not directly measurable, but it can be generally obtained using the normalization of the number of actual peak channel in the energy spectrum divided by the number of saturated ones. The BiSeI-based spectrometer can precisely detect the Am-241 α-ray (5485.56 keV) at room temperature and vacuum environment (0.1 Pa) under varying bias voltages from 6 to 60 V as shown in Figure S7b. It is possible that due to the strong penetrating power of gamma rays, the gamma signals of Am-241 and Co-60 were not detected. The optimal energy resolution of 17.6% could be achieved and the α-ray spectra shift towards higher channel numbers as the applied voltage increases, indicating the improved charge collection efficiency. Figure 2c shows the fitted charge collection efficiency as a function of bias voltage based on the Hecht model; the $\mu\tau$ product of BiSeI is calculated as $1.7\times10^{-4}$ cm$^2$ V$^{-1}$. This value far exceeds that of a-Se[31], the material currently used in commercial imagers, and is comparable to those of some emerging perovskite materials[32]. Figure S7a shows the device structure used for the testing.

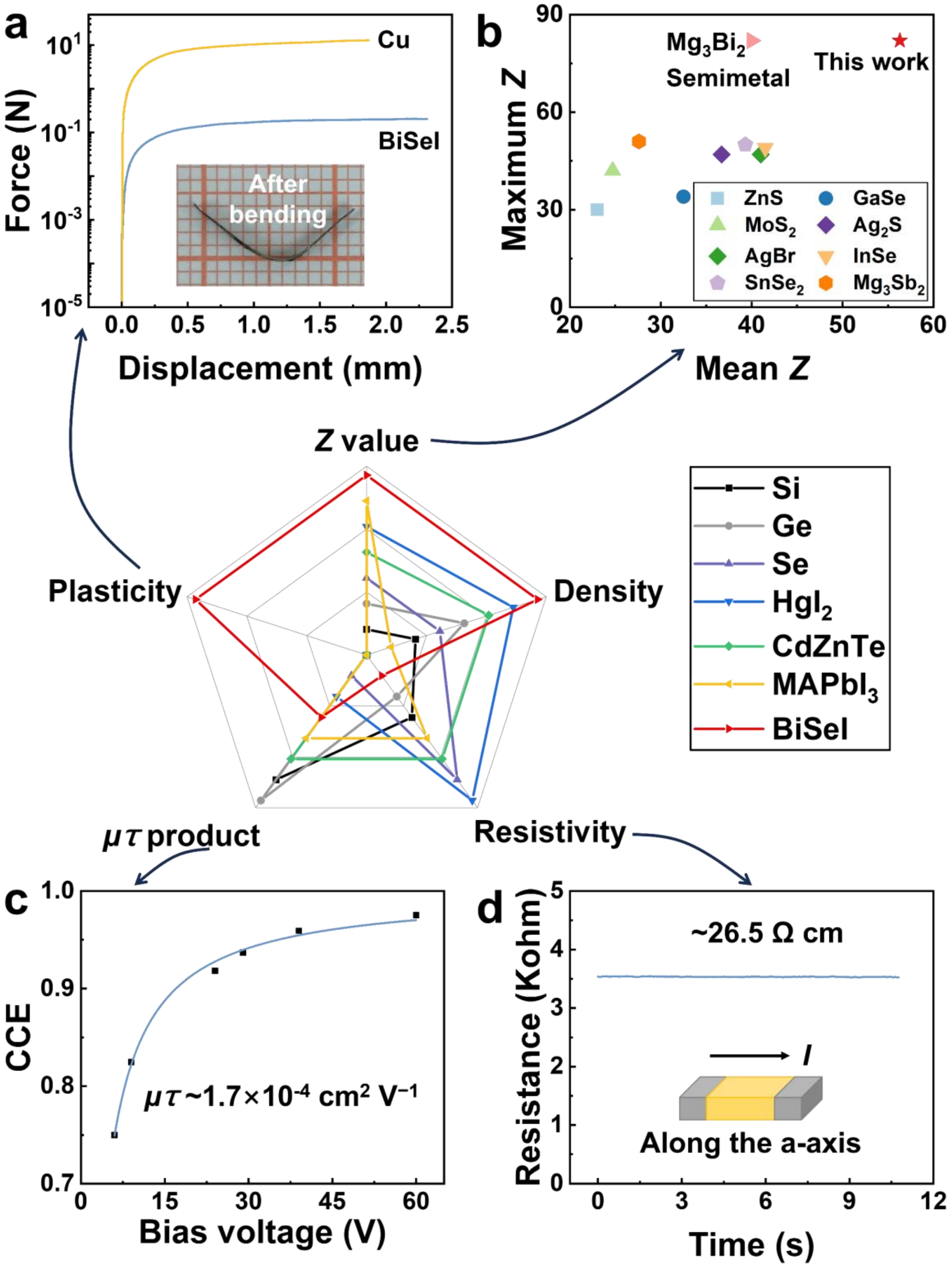

**Figure 2.** Mechanical and electronic properties of BiSeI single crystal. The central radar chart is a comparison of the key figures of merit (*Z* value, $\mu\tau$ product, resistivity, density, and plasticity) among BiSeI and classical X-ray detection materials (Si, HPGe, a-Se, $HgI_2$, CdZnTe, and $MAPbI_3$ perovskite). Please note that mechanical properties refer to the bulk state; flexibility can be achieved in brittle materials through thinning or film fabrication. a) Force-displacement curves from bending test performed on the BiSeI and polycrystalline copper (Cu) at room temperature. The inset is the optical image of deformed BiSeI single crystal after three point bending test. b) Comparison of the atomic number (*Z*

value) of plastic inorganic materials, where the horizontal and vertical coordinates represent the mean and maximum atomic numbers, respectively. c) The bias dependence of the charge collection efficiency (CCE) generated by Am-241 α-ray radiation; the blue line indicates a fit with the Hecht model showing a high $\mu\tau$ product of $1.7\times10^{-4}$ $cm^2$ $V^{-1}$. d) The resistivity of BiSeI measured by the direct current (DC) resistance method. The inset shows the direction of the material tested, where the gold and gray blocks indicate BiSeI and silver electrodes, respectively. The effective sample dimension for resistivity test is $10\times3\times0.25$ $mm^3$.

Another figure of merit for radiation detection materials is resistivity. It is widely accepted that higher resistivity generally correlates with better performance metrics, such as low dark current. BiSeI's resistivity along the a-axis, determined via direct current (DC) measurements, reaches ~26.5 Ω cm (Figure 2d). The bandgap of the synthesized plastic BiSeI single crystal is 1.34 eV determined, by the photoluminescence (PL) spectroscopy (Figure S8). The discrepancy between the theoretical value (1.5~1.6 eV)[33,34] and the measured value may be due to the absence of thermal effects[34], such as lattice expansion and electron-phonon coupling, during the calculation. In general, wide band-gap semiconductors are beneficial for enhancing radiation hardness and resistivity, but not conducive to reducing electron-hole pairs (EHPs) creation energy.[7] At present, many studies[35,36] are focused on increasing the band-gap, while lack the attention to the EHPs creation energy of the radiation materials. This leads to that most reported X-ray detectors exhibit low response current and high energy consumption. Therefore, a medium bandgap material may address the trade-off between the EHPs creation energy and the capability of radiation hardness. Moreover, the resistivity of BiSeI could be further enhanced by doping in the future. For example, the doping of lower-period chalcogens (O, S) or halogens (F, Cl, Br) could increase the bandgap of BiSeI.[37] Another strategy is that the radiation detectors can be based on heterojunctions of BiSeI with wide bandgap materials.[38] This design aims to achieve high carrier mobility while effectively suppressing dark current. Our work adopts this strategy to prepare high-performance X-ray detectors. The overall resistivity of the constructed BiSeI/NiO heterojunction is approximately $1.79\times10^4$ Ω cm, nearly 675 times of the original BiSeI.

When evaluated comprehensively, the overall performance profile of BiSeI is highly competitive against classical detection materials, even though its resistivity remains for improvement. A simple comparison of the $Z$ value, $\mu\tau$ product, resistivity, density, and plasticity of BiSeI with other classical radiation detection materials (including Si, HPGe, a-Se, $HgI_2$, CdZnTe, and $MAPbI_3$ perovskite) is presented in the central radar chart (Figure 2) and Table S1. The $\mu\tau$ product of BiSeI is intermediate compared to those of the classical materials, while its resistivity still needs to be increased. Nevertheless, BiSeI possesses the highest $Z$ value, density and plastic/flexible characteristics, far surpassing other radiation detection materials. The excellent flexibility and semiconducting properties of BiSeI make it particularly suitable for portable and deformable radiation detectors.

## 2.2. The flexible optoelectronic X-ray synapse

The BiSeI/NiO-$V_{Ni}$ heterojunction devices exhibit remarkable X-ray–induced synaptic behavior that closely mimics the responses of biological synapses. A schematic analogy between a biological synapse and the BiSeI/NiO-$V_{Ni}$ double-ended X-ray optoelectronic synapse is presented in **Figure 3**a, and corresponding device photographs are shown in Figure 3b. In a biological synapse, neurotransmitters released by the presynaptic membrane modulate the postsynaptic response. Analogously, in this device, the source electrode (Ag) and leakage electrode (Au) act as the presynaptic and postsynaptic membranes, respectively; the X-ray pulse serves as the action potential; and the BiSeI/NiO-$V_{Ni}$ heterostructure functions as the neurotransmitter medium. The X-ray–induced increase in photocurrent corresponds to the excitatory postsynaptic current (EPSC).

This optoelectronic synapse successfully reproduces synaptic plasticity, a critical feature of biological synapses encompassing both short-term (STP) and long-term plasticity (LTP). STP is associated with short-term memory (STM) and typically lasts for tens of milliseconds to tens of seconds.[39] In contrast, LTP, which corresponds to long-term memory (LTM) persisting for over 100 seconds,[39] refers to the enhanced connectivity between neurons that makes synaptic transmission more effective.

A key manifestation of STP is paired-pulse facilitation (PPF), a phenomenon where a preceding stimulus enhances a subsequent response. To experimentally characterize this behavior, we delivered two consecutive X-ray pulses (1 s interval, 5 s pulse width, 1357 nGy $s^{-1}$ dose rate) and clearly observed the resulting EPSC (Figure S9a). The second EPSC ($P_2$) is greater than the first ($P_1$) because the photogenerated carriers from the first pulse have not fully recombined before the second pulse arrives. The relationship between the paired-pulse facilitation (PPF) index ($P_2/P_1 \times 100\%$) and pulse interval follows a bi-exponential decay (Figure S9b and Equation 4[40]):

$$PPF \text{ index} = A_0 + A_1 \exp(-\Delta t/\tau_1) + A_2 \exp(-\Delta t/\tau_2) \, , \quad (4)$$

Here, $\Delta t$ is the time interval between the two spikes, $A_1$ and $A_2$ are the initial magnitudes of facilitation, $\tau_1$ and $\tau_2$ respectively represent the relaxation times of the rapid and slow phases. In this study, $\tau_1$=0.08 s and $\tau_2$=5.66 s, with $\tau_2$ being more than one order of magnitude larger than $\tau_1$, which agrees with the behavior of biological synapses.[41,42]

The observation of PPF confirms the device's capability for STP. To fully emulate synaptic functionality, however, the X-ray optoelectronic synapses must also exhibit LTP and the ability to transition between STP and LTP states, typically achieved by modulating stimulation. Therefore, we systematically investigated the synaptic plasticity of the BiSeI/NiO-$V_{Ni}$ X-ray optoelectronic synapse by varying X-ray pulses, including pulse number, pulse width, and pulse intensity.

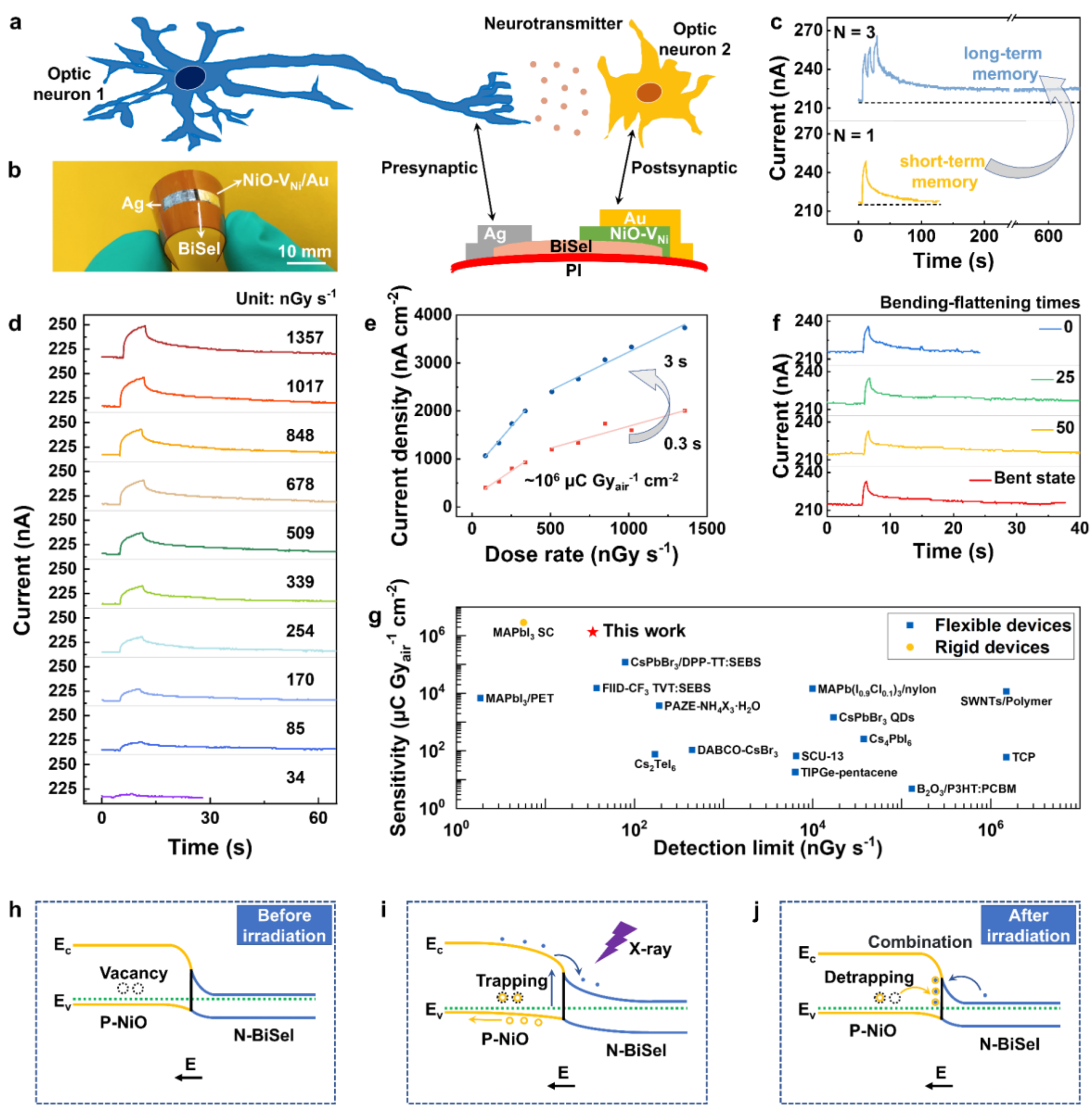

**Figure 3.** Performance of the BiSeI/NiO-$V_{Ni}$ X-ray optoelectronic synapses. a) Schematic analogy of the information transmission related to the visual nerves and the BiSeI/NiO-$V_{Ni}$ X-ray optoelectronic synapses. b) The optical image of a BiSeI/NiO-$V_{Ni}$ X-ray optoelectronic synapse. c) Short-term memory (STM) and long-term memory (LTM) are controlled by the irradiation pulses with different pulse numbers under 1357 nGy $s^{-1}$ at 0.05 V. d) The photocurrent of BiSeI/NiO-$V_{Ni}$ X-ray optoelectronic synapses under the irradiation with different dose rates at 0.05 V for 5 s. e) The photocurrent density of BiSeI/NiO-$V_{Ni}$ X-ray optoelectronic synapses induced by the X-ray irradiation for 0.3 and 3 s as a function of dose rate. f) The photocurrent response of the BiSeI/NiO-$V_{Ni}$ optoelectronic synapse under mechanical bending. The device was tested under a curvature radius of 5 mm after 0, 25, and 50 bending-flattening cycles, or maintained in a bent state, in response to a 0.8 s X-ray pulse (dose rate: 1357 nGy $s^{-1}$). g) The sensitivities and detection limits of currently reported X-ray detectors and the BiSeI/NiO-$V_{Ni}$ X-ray optoelectronic synapses in this work. Energy band diagrams of the BiSeI/NiO-$V_{Ni}$ X-ray optoelectronic synapses h) before X-ray irradiation, i) under X-ray irradiation, and j) after X-ray irradiation.

We first examined the modulation by the number of irradiation pulses. Figure 3c presents the EPSC induced by one-pulse and three-pulse radiations (pulse width of 5 s and time interval of 5 s) at 1357 nGy $s^{-1}$. The memory retention time has been increased from about 100 s to 650 s by increasing the number of pulses, which demonstrates the transition from STP to LTP in the device.

The pulse width (irradiation duration) is also a key factor for modulating synaptic behavior (Figure S10a). The EPSC increases with the pulse width, which is consistent with previous studies[40] on the infrared/visible/ultraviolet synapses. Under continuous irradiation at 1357 nGy $s^{-1}$, extending the irradiation from 0.3 to 21.6 s increases the memory retention time from tens to hundreds of seconds, demonstrating the ability of the device to emulate the synaptic plasticity. To quantify the influence of pulse width on EPSC, we calculated ΔEPSC as shown in Figure S10b. Here, the ΔEPSC representing the change of synaptic weight is defined as the ratio of $(I_{irrad}-I_{dark})/I_{dark}$, where $I_{dark}$ is the current without irradiation and $I_{irrad}$ is the photocurrent during irradiation. The results show a positive correlation between ΔEPSC and irradiation duration. Specifically, the ΔEPSC grows from 4.25% to 21.2% with increasing pulse width and gradually saturates. This trend suggests that the connectivity between synapses strengthens with stimulation duration, and that longer stimulation durations contribute to enhanced memory retention, revealing the tunability of synaptic weight and the realization of LTP. The saturation behavior can be attributed to the dynamic balance between the generation and recombination of photo-generated electron-hole pairs under prolonged illumination. The experimental data are well-fitted by the equation[40]:

$$\Delta EPSC = y_0 + y_1 \exp(-w/t) , \tag{5}$$

where $y_0$ represents the resting facilitation amplitude, $y_1$ represents the initial facilitation magnitude during training, $w$ is the duration of presynaptic spike, and $t$ denotes the facilitation magnitude of presynaptic spikes. The performance of BiSeI/NiO-$V_{Ni}$ optoelectronic synapse displays the basic synaptic behavior as well as the visual models of human memory and forgetting, highlighting its application potential for flexible X-ray synapses.

Figure 3d shows the photocurrents of device induced by X-ray pulses (pulse width of 5 s) with different dose rates. As the dose rate increases from 34 to 1357 nGy $s^{-1}$, more charge carriers are generated, resulting in an enhanced EPSC, following the consistent pattern of our device under different stimulation conditions. This dose rate dependent modulation effectively drives a transition from STP to LTP, further demonstrating the capability of the BiSeI/NiO-$V_{Ni}$ synapse to emulate fundamental neurosynaptic functions under X-ray irradiation.

Particularly, the BiSeI/NiO-$V_{Ni}$ synapses exhibited an ultralow detection limit of 34 nGy $s^{-1}$, about 1/1288 of the dose rate of commercial medical chest X-ray diagnosis.[43] And the applied regulatory voltage ($V_{DS}$ = 0.05 V) is only half of the typical voltage in visual nerve,[44] which is advantageous in terms of power consumption. The power consumption ($E$) can be quantified by the following equation[45]:

$$E = I \times V \times t , \tag{6}$$

where $I$ is the difference between the photocurrent and the dark current, $V$ is the operating voltage, and $t$ is the duration of irradiation pulse (pulse width). A smaller applied voltage results in lower power consumption. Though unable to reach the power consumption level of biological synapses (1~100 fJ)[46], the X-ray optoelectronic synapse proposed in this study (power consumption per spike is ~90 pJ, 0.2 s irradiation at 1357 nGy $s^{-1}$, Figure S11b) is at the leading level among the devices reported to date[3,7,8] (Figure S11a).

Moreover, the sensitivity of the BiSeI/NiO-$V_{Ni}$ optoelectronic synapse is also related to the irradiation duration and dose rate, which is similar to the conclusion in previous literature[7]. Unlike the classic X-ray photoelectric detectors, the photocurrent gain ($I_{irrad}$-$I_{dark}$) of the X-ray optoelectronic synapse is increased with the irradiation time. Therefore, when defining the X-ray induced current density and calculating the sensitivity, it is necessary to study the photocurrent gain of the X-ray optoelectronic synapse for different irradiation times. As shown in Figure 3e, the data extracted from Figure 3d under 0.3 s irradiation shows that the X-ray induced current density is nonlinearly related to the dose rate. Using 500 nGy $s^{-1}$ as the dividing point, the slopes of the fitted curves represent the sensitivity of detector in the low and high dose rate ranges, respectively. It is evident that the detector's sensitivity within the 0 to 500 nGy $s^{-1}$ range is higher than within the 500 to 1500 nGy $s^{-1}$ range, but in both ranges, it surpasses most flexible devices and even rivals the performance of some perovskite single crystals. The detailed comparisons are provided in Figure 3g and Table S2. The data extracted from Figure 3d under 3 s irradiation also presents a nonlinear response. However, prolonged irradiation enhances the detector's sensitivity, with values rising from $2.2 \times 10^6$ μC $Gy_{air}^{-1}$ $cm^{-2}$ to $3.8 \times 10^6$ μC $Gy_{air}^{-1}$ $cm^{-2}$ in the lower dose rate range (0 to 500 nGy $s^{-1}$) and from $0.91 \times 10^6$ μC $Gy_{air}^{-1}$ $cm^{-2}$ to $1.6 \times 10^6$ μC $Gy_{air}^{-1}$ $cm^{-2}$ in the higher dose rate range (500-1500 nGy $s^{-1}$), respectively. Clearly, the sensitivity is changed by different irradiation times. Under short-term irradiation, the defects in the synapse material trap charge carriers, decreasing the sensitivity. As irradiation time increases, these traps are gradually filled, allowing subsequently generated carriers to be more efficiently collected by the electrodes. Consequently, the sensitivity of the BiSeI/NiO-$V_{Ni}$ X-ray synapse improves with prolonged irradiation. The influence of the dose rate ranges on detector sensitivity is also related to carrier transport behavior. In the high dose rate regime, the BiSeI/NiO-$V_{Ni}$ heterojunction generates a significantly increased number of electron-hole pairs upon photon absorption. The excessively high carrier concentration leads to accelerated recombination rates, resulting in weakened effective signals. Consequently, the device exhibits lower sensitivity in the high dose rate range.

Beyond the irradiation parameters influence characteristics, we verified the bending stability of the device at a curvature radius of 5 mm. Under an X-ray pulse with a dose rate of 1357 nGy $s^{-1}$ and an irradiation time of 0.8 s, the photocurrent variation over time of the BiSeI/NiO-$V_{Ni}$ optoelectronic synapse after 0, 25, and 50 cycles of bending and flattening, as well as when maintained in bent state, is shown in Figure 3f. The results revealed a moderate photocurrent degradation of approximately 10% after 50 bending cycles, which is primarily attributed to the formation of dislocation defects as investigated in earlier sections. This level of degradation is comparable to the ~10% change in electrical resistance reported for plastic InSe under a similar number of bending cycles, while being achieved under a more aggressive bending condition (InSe at 8.25 mm curvature radius).[47] Notably, the increase in dislocation defects induced by bending directly leads to a prolonged carrier recombination time, therefore presenting a viable strategy for enhancing memory retention. While further study is required to minimize the optoelectronic performance loss, these results unequivocally demonstrate the fundamental feasibility of our device for applications requiring moderate flexibility.

To elucidate the origin of these synaptic behaviors, the band alignment at the BiSeI/NiO-$V_{Ni}$ interface was systematically characterized. The XPS survey (Figure S12a) reveals that the atomic ratio of Bi, Se, and I in BiSeI is approximately 9.3:9.5:10, which closely aligns with the theoretical 1:1:1 ratio as the experimental error is considered. This result implies the high crystal quality of BiSeI. Due to the superior crystallinity of BiSeI single crystals, the photodetectors based on the device structure of metal/BiSeI/metal do not exhibit significant memory phenomenon.[33] To introduce the memory functionality, a 50 nm-thick layer of NiO-$V_{Ni}$ was deposited onto the BiSeI single crystal using electron beam evaporation. NiO is a P-type material. Therefore, a P-N junction is formed with N-type BiSeI. The atomic ratio of Ni and O in the evaporated NiO layer is 6.9:10 (Figure S12b), significantly deviating from the stoichiometric 1:1 ratio, indicating the existence of a substantial number of Ni vacancies. Raman spectroscopy further supports the presence of Ni vacancies. As shown in Figure S13, the Raman spectrum of electron beam-evaporated NiO film (yellow) differs significantly from that of commercial polycrystalline bulk NiO (green). The latter displays one prominent peak near 545 $cm^{-1}$, corresponding to the one-phonon first-order longitudinal-optical (1P LO) vibration.[48] In contrast, the former exhibits a significant attenuation of the 1P LO peak and quenching of other Raman peaks, indicating excess defects caused by the Ni vacancies. These defects disrupt the vibrational modes, reducing the intensity and resolution of Raman peaks. It is the presence of Ni vacancies that underlies the memory behavior in the BiSeI/NiO-$V_{Ni}$ heterojunction optoelectronic synapses.

To analyze the band alignment of the BiSeI/NiO-$V_{Ni}$ heterostructure, the PL (Figure S8) and optical absorption (Figure S14) measurements were conducted. The bandgap of BiSeI was previously determined to be 1.34 eV. For the NiO-$V_{Ni}$ film deposited on a quartz substrate, the UV-visible absorption spectrum shows a cutoff wavelength of 362 nm, corresponding to an optical bandgap of approximately 3.42 eV. The Kelvin probe force microscopy (KPFM, Figure S15) was used to determine the Fermi levels of BiSeI and NiO-$V_{Ni}$. Based on these results and previous literature studies[34,49], the energy band diagram of the BiSeI/NiO-$V_{Ni}$ heterojunction was constructed (Figure S16), revealing a type-II band alignment. This configuration promotes the separation of photogenerated carriers, thereby enhancing the photoelectric response.

Before the X-ray irradiation (Figure 3h), the Fermi levels of BiSeI and NiO-$V_{Ni}$ align at the P-N junction, resulting in the band bending. The blue and yellow regions represent BiSeI and NiO-$V_{Ni}$, respectively. Due to the diffusion of free carriers, a built-in electric field is formed from BiSeI to NiO-$V_{Ni}$, which is defined as the initial state. During the X-ray irradiation, the BiSeI/NiO-$V_{Ni}$ active layer absorbs photons to generate excitons, which diffuse towards the P-N junction. The excitons in NiO-$V_{Ni}$ transfer electrons to BiSeI, while the excitons in BiSeI transfer holes to NiO-$V_{Ni}$, enabling effective charge separation. The electrons and holes migrate to the negative and positive electrodes through BiSeI and NiO-$V_{Ni}$, respectively, generating a photocurrent. This process flattens the energy band, compared to that before irradiation (Figure 3i).

As the irradiation continues, many photogenerated carriers accumulate at the BiSeI/NiO-$V_{Ni}$ interface. Some photogenerated holes are captured by the Ni vacancies, causing the EPSC to increase with prolonged irradiation. After the irradiation ceased (Figure 3j), the photogenerated electrons begin to recombine with holes, reducing the EPSC. However, the holes trapped by Ni vacancies delay the recombination, resulting in a gradual decrease of current and a bent energy band. This persistent current demonstrates the memory function of BiSeI/NiO-$V_{Ni}$ heterojunction, mimicking the optoelectronic synapse behavior.

We also demonstrate the potential of the flexible X-ray optoelectronic synapse in practical medical imaging applications. In advanced X-ray medical equipment (**Figure 4**a), detectors are often designed with specific curvatures to minimize optical distortions and improve imaging qualities—a feature that highlights the advantage of our flexible devices. As illustrated in Figure 4b, a curved detector geometry is effectively realized with a bent 1×6 synapse array, which highlights its conformal mounting capability. The six devices exhibit highly consistent photoresponse under X-ray irradiation (Figure 4e), indicating good spatial uniformity.

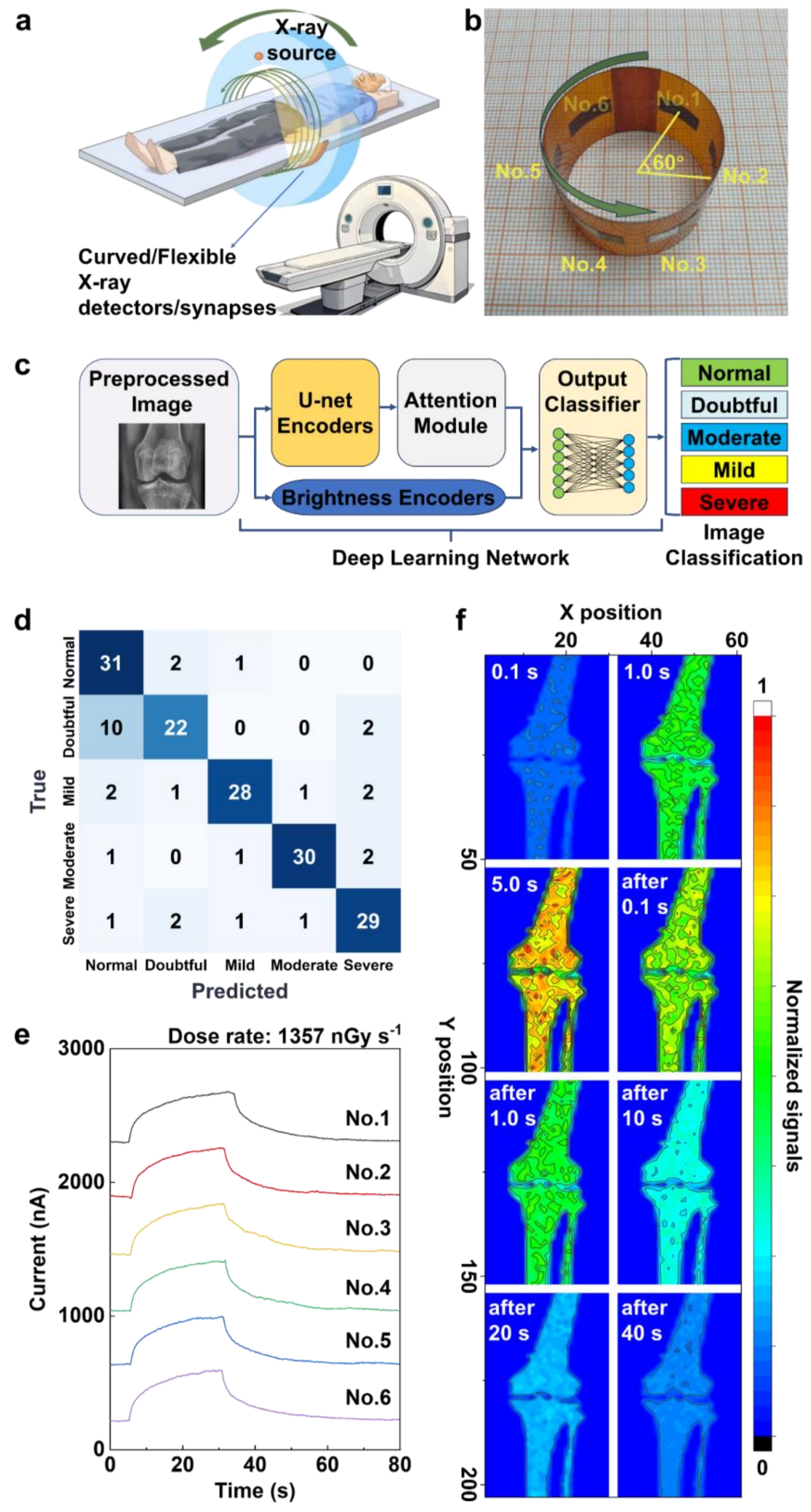

**Figure 4.** Application of flexible BiSeI/NiO-$V_{Ni}$ X-ray optoelectronic synapses in medical machine vision. a) Schematic diagram of the flexible/curved radiation detectors used for medical imaging. b) Optical photograph of a BiSeI/NiO-$V_{Ni}$ X-ray optoelectronic synapse array, featuring equidistantly positioned 1×6 pixels on a polyimide (PI) film, where the round BiSeI/NiO-$V_{Ni}$ X-ray optoelectronic synapse array has a curvature radius of ~16 mm. c) Deep learning model for orthopedic X-ray image diagnosis. d) Confusion matrix of the trained neural network in multi-class (different arthritis severity levels)

classification tasks. e) Photocurrent uniformity across the round synapse array. The device was exposed to an X-ray pulse with a dose rate of 1357 nGy $s^{-1}$, with the source positioned 5 cm from each synapse. f) Neuromorphic evolution of a "knee joint" image during synaptic learning and forgetting processes. The image clarity improves with increased X-ray irradiation time (0.1 s, 1 s, and 5 s at 1357 nGy $s^{-1}$) during the learning phase and gradually decays while remaining distinguishable during the forgetting phase (0.1 s, 1 s, 10 s, 20 s, and 40 s after irradiation cessation), demonstrating persistent memory retention.

Moreover, the proposed optoelectronic synapses perform imaging functions that are comparable to conventional CMOS and CCD sensors, while enabling neuromorphic image enhancement and learning through their synaptic characteristics. This capability can be integrated with deep learning techniques to assist in orthopedic X-ray diagnosis. For example, osteoarthritis diagnosis and grading heavily rely on radiographic interpretation, which remains challenging even for experienced radiologists. To verify this potential, we constructed a convolutional neural network (CNN) that mimics the operations of the BiSeI/NiO-$V_{Ni}$ X-ray optoelectronic synapse to identify and grade arthritic bone images (Figure 4c and Figure S17). After training, the model achieved an overall classification accuracy of 82.35% across multiple arthritis levels (Figure 4d). When simplified to a binary classification task (arthritis vs. non-arthritis), the accuracy reached 90% (153/170), demonstrating promising diagnostic support potential.

The neuromorphic preprocessing capability of the BiSeI/NiO-$V_{Ni}$ synapse underpins this performance. Figure 4f illustrates the evolution of a "knee joint" image during learning and forgetting processes. As the learning time (irradiation duration) increases, the image becomes clearer due to enhanced synaptic conductance. Under the stimulation of a 5 s X-ray pulse (1357 nGy $s^{-1}$), the pattern remains clearly distinguishable even after the source is turned off. Over time (0.1 s to 40 s), the image gradually blurs, yet a 18.2% retention level persists at 50 s, reflecting a prolonged memory effect. This behavior mimics human visual memory—the longer an object is observed, the more clearly its features are retained. Such characteristics are beneficial for implementing bio-inspired vision systems in next-generation X-ray imaging and diagnostic technologies.

## 3. Conclusion

We developed a quasi-1D plastic inorganic semiconductor, BiSeI, that outperforms the conventional materials such as Si, Ge, and CdZnTe in terms of photon absorption, allowing for thinner and lighter X-ray detectors. A flexible X-ray optoelectronic synapse is constructed based on the heterojunction of single crystalline BiSeI and polycrystalline NiO-$V_{Ni}$, in which the mimetic synaptic plasticity of the NiO-$V_{Ni}$ empowers the device with memory abilities. The devices exhibit PPF and transition from STP to LTP, while maintaining excellent mechanical flexibility and stability. The synaptic behavior of BiSeI/NiO-$V_{Ni}$ devices can be tuned by adjusting the irradiation parameters to simulate STM/LTM, offering new pathways for the bio-inspired radiation detection systems. These devices can imitate the learning and adaptive capabilities of the human synapses, which is crucial for improving the accuracy and sensitivity of X-ray detection in various applications. The power consumption and limit of detection of the proposed device is also significantly lower than the other X-ray synapses reported to date.

In summary, this work highlights the potential of BiSeI/NiO-$V_{Ni}$ as a leading material for the next-generation flexible and efficient X-ray detectors. The unique combination of high photoelectric absorption coefficient, excellent flexibility, and synaptic memory capability sets BiSeI/NiO-$V_{Ni}$ device apart from traditional detectors, overcoming the current challenges in the industry of X-ray detection. The advancements made in this study open up exciting possibilities for the flexible, low-power, and high-performance X-ray detection technologies, paving the way for innovative applications in medical, environmental, and industrial fields.

## 4. Experimental Methods

*Characterization:*

The initial sample dimension for three points bending is $15\times3\times0.25$ mm$^3$, using Instron-5969 universal testing machine with a constant loading rate of 0.05 mm min$^{-1}$. Nanoindentation data were collected on Agilent G200. The scanning electron microscope (SEM) images were carried out on ZEISS Merlin Compact. The transmission electron microscopy (TEM) and spherical aberration corrected transmission electron microscopy (AC-TEM) data of BiSeI were obtained using the Talos F200X and JEM-ARM200F NEOARM. The TEM samples were done by a focused ion beam (FIB, FEI Helios NanoLab 600i DualBeam). Optical photographs were taken by VIVO IQOO Z7I. Resistivity test was recorded by Keithley 2400 using a dc resistance method. The Roman spectra and PL spectrum was obtained by using Renishaw inVia Raman microscope with a 532 nm excitation wavelength at room temperature. The BiSeI crystal was examined by X-ray diffraction using Cu Kα Source (Panalytical Instruments, X 'PERT). The absorption coefficients were calculated using the NIST Photon Cross Section Database (XCOM). The X-rays tube was manufactured by Source-Ray, Inc (Model SB-80-500). The X-ray source operated at an accelerating voltage of 50 kV, and the dose rate was adjusted by changing the tube current (10 μA–400 μA). The α-ray spectroscopies were operated coupled with ORTEC system (preamplifier 142B, main amplifier 572A and multichannel pulse analyzer EASY-MCA), generated by 241Am α-ray radiation (5485.56 keV, activity 0.06 MBq). The Kelvin probe force microscopy (KPFM) dates were tested by Dimension FastScan® Atomic force microscope (AFM). The XPS surveys were revealed by X-ray photoelectron spectroscopy (XPS) ESCALAB 250Xi. The UV-visible absorption spectrum was determined by PerkinElmer Lambda 950 ultraviolet-visible spectrophotometer.

*Device Fabrication and Measurement:*

High-purity Au (4N, 99.99%), Ag (4N, 99.99%), Bi (5N, 99.999%), Se (5N, 99.999%), I (4N, 99.99%), NiO (3N, 99.9%) and $MoO_3$ (4N, 99.99%) were obtained from ZhongNuo Advanced Material (Beijing) Technology Co., Ltd, and were used as the raw materials. The structure of the X-ray optoelectronic synapses was Au/NiO-$V_{Ni}$/BiSeI/Ag, where BiSeI crystal was obtained by a CVT process as described in the main text and had a thickness of 0.25 mm. The NiO-$V_{Ni}$ film (50 nm) was prepared by electron beam evaporation. The Au and Ag electrodes were deposited by thermal evaporation with a thickness of 100 nm, and the Au/Ag electrodes were coated with conductive graphite and silver adhesives, respectively, for follow-up tests. The channel length and width of the device with charge transport along the a-axis were 1 and 3 mm respectively. The entire device was attached to a flexible polyimide (PI) or polyethylene terephthalate (PET) substrate by 3M™ 100MP high performance acrylic adhesive (F9460PC). To ensure the accuracy of the data, a tungsten shield with a 1×3 $mm^2$ slit was placed between the single device and the X-ray source. The distance between the source and the device was 5 cm. The 1×6 synapse array shared an identical device structure with the single device, and the photocurrent uniformity was characterized under the same conditions. The single-pixel imaging experiments were carried out on the translation stage to control the two-dimensional (X–Y) movement of the imaging object (a stainless mask featuring a knee joint pattern, measuring 50 × 70 $mm^2$, Figure S18). The actual imaging area was the central 30 × 50 $mm^2$ region of the mask, with a 10 mm peripheral border reserved for mounting and for ensuring pattern integrity. The distance between the imaging object and the synapse was 0.5 mm. To facilitate scanning and enhance resolution, the device's channel size was reduced to 1 × 1 $mm^2$ while retaining all other structural parameters. The response current was recorded using the Keithley 2400 source measure unit. Sampling coordinates were determined by moving the mask in 1 mm increments along both the X and Y axes, respectively. The pixel was illuminated and measured one by one, and the pixel current was obtained sequentially. All the measurements were performed at the atmospheric pressure and room temperature.

The structure of the α-ray spectrometer was Au/Cr/$MoO_3$/BiSeI/Ag, where the distance between the Cr/Au electrodes (10/100 nm) and the Ag electrode (100 nm) was 250 μm. $MoO_3$ (50 nm) acted as a buffer layer, which was evaporated by electron beam. A fixed 0.5 mm separation, maintained by a silicon wafer between the α-source and the spectrometer, maximized the count rate and minimized energy loss straggling in the residual atmosphere.

*Convolutional Neural Network:*

This study utilized a retrospective dataset of knee joint X-ray images from patients with arthritis, which was provided by the Department of Orthopedics at the First Affiliated Hospital of Harbin Medical University. Informed consent was obtained from all individual participants included in the study at the time of data collection. To protect patient privacy, all personally identifiable information was removed from the images prior to analysis. The model was developed using a dataset of 850 orthopedic X-ray images, with 170 samples for each of five clinical labels. These labels categorized the severity of arthritis, including 'Normal', 'Doubtful', 'Mild', 'Moderate', and 'Severe'. Each label in the dataset was randomly partitioned, with 80% of the samples (136 images) allocated as the training set and the remaining 20% (34 images) reserved as the validation set. Each sample image was processed by the identical preprocessing protocol. First, size normalization and random horizontal flipping were applied. Subsequently, the image enhancement technology based on contrast-limited adaptive histogram equalization (CLAHE) was employed to improve local feature representation, followed by the application of gamma correction algorithm to perform non-linear mapping of the illumination distribution, thereby sharpening detail discernibility in low-illumination areas. The resulting preprocessed images were obtained. Then, the classification task was completed by a deep learning model based on U-Net encoders and an attention module. The input to the network was the preprocessed image sample. Features were extracted from the sample by U-Net encoders and an attention module. The U-Net encoders features were transformed into a spatial weight map by 1×1 convolution, and the pixel-level mask was generated by Sigmoid activation. The weighted features significantly strengthened the response of the target areas. In addition, the network contained a set of brightness encoders to capture the local brightness differences reflected in the X-ray images associated with different levels of arthritis severity. Finally, the fused feature map, obtained by concatenating the two parts, was fed into a fully connected layer with a 512-dimensional hidden layer to make the final classification. After training 100 Epochs, we got the classification results presented in the confusion matrix.

*Declaration:*

During the preparation of this manuscript, the authors used ChatGPT solely for the purpose of improving language fluency and grammar in the initial draft. All scientific content, data analysis, interpretation, and conclusions are entirely the product of the authors' work. The AI tool was not used to generate any scientific ideas, data, or interpretations, and it is not listed as an author.

**Supporting Information**

Supporting Information is available from the Wiley Online Library or from the author.

**Acknowledgements**

This work was supported by the National Natural Science Foundation of China (NSFC, No.52372042), National Key R&D Program of China (2019YFA0705201), Foundation for Innovative Research Groups of the National Natural Science Foundation of China (No. 51521003), Self-Planned Task of State Key Laboratory of Robotics and System (HIT) (No. SKLRS202212B), the National Natural Science Foundation of China (52405585), the Postdoctoral Fellowship Program of CPSF under Grant Number GZC20242223.

**Conflict of Interest**

The authors declare no conflict of interest.

References

[1] P. J. Withers, C. Bouman, S. Carmignato, V. Cnudde, D. Grimaldi, C. K. Hagen, E. Maire, M. Manley, A. Du Plessis, S. R. Stock, *Nature Reviews Methods Primers* **2021**, *1*, 18.

[2] M. Yabashi, H. Tanaka, *Nature Publishing Group* **2011**, DOI 10.1038/nphoton.2016.251.

[3] H. Yin, J. Pang, S. Zhao, H. Wu, Z. Song, X. Li, Z. Zheng, L. Xu, J. Tang, G. Niu, *Innovation* **2024**, *5*, 100654.

[4] P. Barba, J. Stramiello, E. K. Funk, F. Richter, M. C. Yip, R. K. Orosco, *Surgical Endoscopy* **2022**, *36*, 2771.

[5] K. Nagatani, S. Kiribayashi, Y. Okada, K. Otake, K. Yoshida, S. Tadokoro, T. Nishimura, T. Yoshida, E. Koyanagi, **2013**, *30*, 44.

[6] Y. Liu, S. Yu, Z. Zhang, X. Hou, M. Ding, X. Zhao, G. Xu, X. Zhou, S. Long, **2024**, DOI 10.1021/acsami.4c01255.

[7] H. Liang, X. Tang, H. Shao, R. Zhu, S. Deng, X. Zhan, T. Zhu, J. Wang, J. Zhang, G. Zhang, Z. Mei, *Advanced Science* **2024**, *11*, 1.

[8] Y. Liu, S. Yu, Z. Zhang, X. Hou, M. Ding, X. Zhao, G. Xu, X. Zhou, S. Long, *ACS Applied Materials and Interfaces* **2024**, *16*, 24871.

[9] X. Ou, X. Qin, B. Huang, J. Zan, Q. Wu, Z. Hong, L. Xie, *Nature* **2021**, *590*, DOI 10.1038/s41586-021-03251-6.

[10] Z. Jin, Q. Qian, O. F. Mohammed, Z. Zang, **2025**, DOI 10.1016/j.matt.2025.102261.

[11] Z. Lyu, P. Fan, T. Xu, R. Wang, Y. Liu, S. Wang, Z. Wu, T. Ma, in *2019 IEEE Nuclear Science Symposium and Medical Imaging Conference (NSS/MIC)*, **2019**, pp. 1–3.

[12] K. Nikolaou, F. Bamberg, A. Laghi, G. D. Rubin, *Multislice CT Fourth Edition*, **2019**.

[13] B. J. Kim, B. Shao, A. T. Hoang, S. Yun, J. Hong, J. Wang, A. K. Katiyar, S. Ji, D. Xu, Y. Chai, J.-H. Ahn, *Nature Electronics* **2025**, *8*, 147.

[14] S. Siebentritt, M. Igalson, C. Persson, S. Lany, *Progress in Photovoltaics: Research and Applications* **2010**, *18*, 390.

[15] Z. Gao, C. Leng, H. Zhao, X. Wei, H. Shi, Z. Xiao, *Advanced Materials* **2024**, *36*, 1.

[16] J. Ghosh, P. J. Sellin, P. K. Giri, *Nanotechnology* **2022**, *33*, DOI 10.1088/1361-6528/ac6884.

[17] B. Xiao, M. Zhu, L. Ji, B. Bin Zhang, J. Dong, J. Yu, Q. Sun, W. Jie, Y. Xu, *Journal of Crystal Growth* **2019**, *517*, 7.

[18] M. Chen, L. Li, M. Xu, W. Li, L. Zheng, X. Wang, *Research* **2023**, *6*, DOI 10.34133/research.0066.

[19] X. Wen, Z. Lu, X. Yang, C. Chen, M. A. Washington, G. C. Wang, J. Tang, Q. Zhao, T. M. Lu, *ACS Applied Materials and Interfaces* **2023**, *15*, 22251.

[20] M. Li, Y. Shen, K. Luo, Q. An, P. Gao, P. Xiao, Y. Zou, *Nature Materials* **2023**, *22*, 958.

[21] L. F. Kourkoutis, M. K. Parker, V. Vaithyanathan, D. G. Schlom, D. A. Muller, *Physical Review B* **2011**, *84*, 75485.

[22] Y. Wang, W. Zhang, *Micron* **2019**, *125*, 102715.

[23] J. Luo, J. Chen, Z. Gao, J. Zhou, J. Zhang, P. Qiu, S. Shen, S. Dong, L. Chen, X. Shi, *Advanced Materials* **2025**, *14083*, 1.

[24] M. A. Reyes-Martinez, A. L. Abdelhady, M. I. Saidaminov, D. Y. Chung, O. M. Bakr, M. G. Kanatzidis, W. O. Soboyejo, Y. L. Loo, *Advanced Materials* **2017**, *29*, 1.

[25] Z. Gao, T. R. Wei, T. Deng, P. Qiu, W. Xu, Y. Wang, L. Chen, X. Shi, *Nature Communications* **2022**, *13*, DOI 10.1038/s41467-022-35229-x.

[26] A. A. Balandin, F. Kargar, T. T. Salguero, R. K. Lake, *Materials Today* **2022**, *55*, 74.

[27] Q. Chen, J. Wu, X. Ou, B. Huang, J. Almutlaq, A. A. Zhumekenov, X. Guan, S. Han, L. Liang, Z. Yi, J. Li, X. Xie, Y. Wang, Y. Li, D. Fan, D. B. L. Teh, **2018**.

[28] S. Yakunin, D. N. Dirin, Y. Shynkarenko, V. Morad, I. Cherniukh, O. Nazarenko, D. Kreil, T. Nauser, M. V. Kovalenko, *Nature Photonics* **2016**, *10*, 585.

[29] K. M. McCall, Z. Liu, G. Trimarchi, C. C. Stoumpos, W. Lin, Y. He, I. Hadar, M. G. Kanatzidis, B. W. Wessels, *ACS Photonics* **2018**, *5*, 3748.

[30] Q. Xu, H. Wei, W. Wei, W. Chuirazzi, D. DeSantis, J. Huang, L. Cao, *Nuclear Instruments and Methods in Physics Research, Section A: Accelerators, Spectrometers, Detectors and Associated Equipment* **2017**, *848*, 106.

[31] S. O. Kasap, *Journal of Physics D: Applied Physics* **2000**, *33*, 2853.

[32] S. Yakunin, M. Sytnyk, D. Kriegner, S. Shrestha, M. Richter, G. J. Matt, H. Azimi, C. J. Brabec, J. Stangl, M. V. Kovalenko, W. Heiss, *Nature Photonics* **2015**, *9*, 444.

[33] Y. Li, S. Wang, J. Hong, N. Zhang, X. Wei, T. Zhu, Y. Zhang, Z. Xu, K. Liu, M. Jiang, H. Xu, *Small* **2023**, *19*, 1.

[34] A. M. Ganose, K. T. Butler, A. Walsh, D. O. Scanlon, *Journal of Materials Chemistry A* **2016**, *4*, 2060.

[35] S. Johnsen, Z. Liu, J. A. Peters, J. H. Song, S. C. Peter, C. D. Malliakas, N. K. Cho, H. Jin, A. J. Freeman, B. W. Wessels, M. G. Kanatzidis, *Chemistry of Materials* **2011**, *23*, 3120.

[36] W. Lin, C. C. Stoumpos, O. Y. Kontsevoi, Z. Liu, Y. He, S. Das, Y. Xu, K. M. McCall, B. W. Wessels, M. G. Kanatzidis, *Journal of the American Chemical Society* **2018**, *140*, 1894.

[37] U. V Ghorpade, M. P. Suryawanshi, M. A. Green, T. Wu, X. Hao, K. M. Ryan, *Chemical Reviews* **2023**, *123*, 327.

[38] H. Hu, W. Zhen, Z. Yue, R. Niu, F. Xu, W. Zhu, K. Jiao, M. Long, C. Xi, W. Zhu, C. Zhang, *Nanoscale Advances* **2023**, *5*, 6210.

[39] H. Shao, Y. Li, W. Yang, X. He, L. Wang, J. Fu, M. Fu, H. Ling, P. Gkoupidenis, F. Yan, L. Xie, W. Huang, *Advanced Materials* **2023**, *35*, 1.

[40] J. Wang, N. Ilyas, Y. Ren, Y. Ji, S. Li, C. Li, F. Liu, D. Gu, K. W. Ang, *Advanced Materials* **2024**, *36*, 1.

[41] R. S. Zucker, W. G. Regehr, **2002**, 355.

[42] S. Liu, J. Guan, L. Yin, L. Zhou, J. Huang, Y. Mu, S. Han, X. Pi, G. Liu, P. Gao, S. Zhou, **2022**, DOI 10.1021/acs.jpclett.2c02900.

[43] Y. Bian, K. Liu, Y. Ran, Y. Li, Y. Gao, Z. Zhao, M. Shao, Y. Liu, J. Kuang, Z. Zhu, M. Qin, Z. Pan, M. Zhu, C. Wang, H. Chen, J. Li, X. Li, Y. Liu, Y. Guo, *Nature Communications* **2022**, *13*, 1.

[44] T. Quintela-López, H. Shiina, D. Attwell, *Current Biology* **2022**, *32*, R650.

[45] Y. Lee, H. L. Park, Y. Kim, T. W. Lee, *Joule* **2021**, *5*, 794.

[46] S. B. Laughlin, R. R. de Ruyter van Steveninck, J. C. Anderson, *Nature Neuroscience* **1998**, *1*, 36.

[47] T. R. Wei, M. Jin, Y. Wang, H. Chen, Z. Gao, K. Zhao, P. Qiu, Z. Shan, J. Jiang, R. Li, L. Chen, J. He, X. Shi, *Science* **2020**, *369*, 542.

[48] N. Bala, H. K. Singh, S. Verma, S. Rath, *Physical Review B* **2020**, *102*, 1.

[49] S. Kossar, R. Amiruddin, A. Rasool, *Microelectronic Engineering* **2022**, *254*, 111669.

A flexible X-ray optoelectronic synapse based on plastic BiSeI/NiO heterojunctions achieves high sensitivity, ultralow detection limits, and memory function, enabling intelligent machine vision for orthopedic diagnosis.

Qiao Wang[+], Pengfei Li[+], Huan Liu[+], Pengyue Zhao, Shuhao Kang, Kunhao Wang, Jinliang Yu, Yajing Wang, Haiying Xiao, Yuqing Wang, Wanqi Liu, Wenxiang Dou, Hsu-Sheng Tsai*, Wenbo Wang*, Jialu Li*, Ping-An Hu*

**Flexible X-ray Optoelectronic Synapses based on Plastic Inorganic Semiconductor BiSeI**

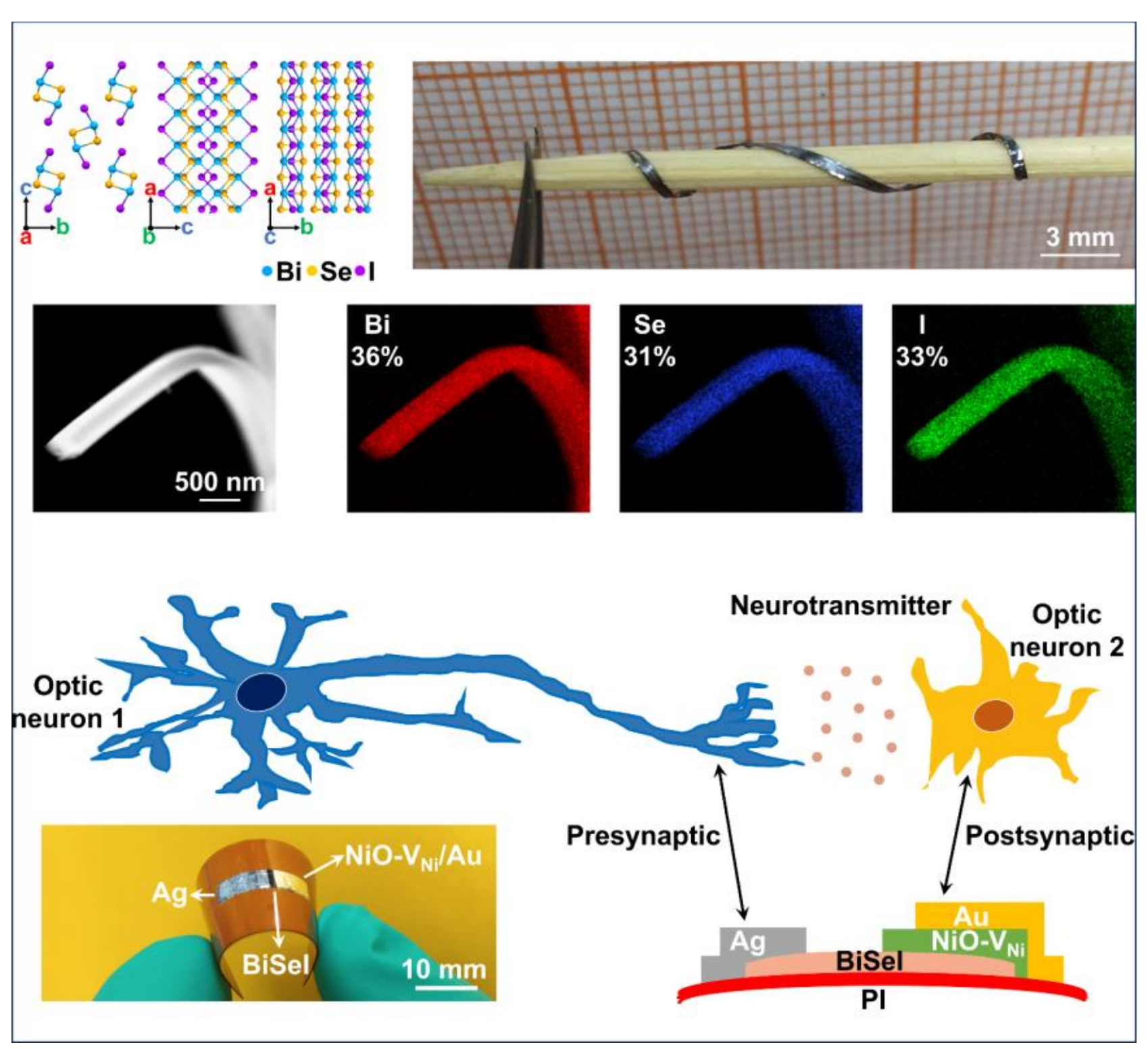